\documentclass[12pt,preprint]{aastex}

\def\saturnring1  {\mbox{\wasyfamily\char21}}

\def\saturnring2 {\mbox{\wasyfamily\char31}}

\shorttitle{FUSE Spectroscopy of BB Dor} 
\shortauthors{Godon et al.}
\begin{document}
\bibliographystyle{apj}

\title{Far Ultraviolet Spectroscopic Explorer Spectroscopy of the
Nova-like BB Doradus
\altaffilmark{1}}

\author{Patrick Godon\altaffilmark{2}, Edward M. Sion} 
\affil{Department of Astronomy and Astrophysics, 
Villanova University,
Villanova, PA 19085}
\email{patrick.godon@villanova.edu} 
\email{edward.sion@villanova.edu}

\author{Paul E. Barrett} 
\affil{United States Naval Observatory, 
Washington, DC 20392}
\email{barrett.paul@usno.navy.mil} 

\author{Paula Szkody}
\affil{Department of Astronomy,
University of Washington,
Seattle, WA 98195}
\email{szkody@astro.washington.edu} 

\and

\author{Eric M. Schlegel}
\affil{Department of Physics and Astronomy, 
University of Texas,
San Antonio, TX 78249} 
\email{eric.schlegel@utsa.edu} 

\altaffiltext{1}
{Based on observations made with the 
NASA-CNES-CSA Far Ultraviolet Spectroscopic
Explorer. {\it{FUSE}} was operated for NASA 
by the Johns Hopkins University under
NASA contract NAS5-32985} 
\altaffiltext{2}
{Visiting at the Space Telescope Science Institute, Baltimore, MD 21218,
godon@stsci.edu}

\begin{abstract}

We present an analysis of the Far Ultraviolet Spectroscopic Explorer
({\it{FUSE}}) spectra of the little-known southern nova-like BB
Doradus.  The spectrum was obtained as part of our Cycle 8 {\it FUSE}
survey of high declination nova-like stars.

The {\it FUSE} spectrum of BB Dor, observed in a high state, is
modeled with an accretion disk with a very low inclination (possibly
lower than 10$^{\circ}$). Assuming an average WD mass of
$0.8M_{\odot}$ leads to a mass accretion rate of $10^{-9}
M_{\odot}$/yr and a distance of the order of $\sim$650 pc, consistent
with the extremely low galactic reddening in its direction.  The
spectrum presents some broad and deep silicon and sulfur absorption
lines, indicating that these elements are over-abundant by 3 and 20
times solar, respectively.

\end{abstract}

\keywords{
accretion, accretion disks
--- 
novae, cataclysmic variables 
--- 
stars: abundances 
--- 
stars: individual (BB Doradus) 
--- 
stars: nova-likes  
--- 
stars: white dwarfs 
}  

\section{Introduction} 

\subsection{Cataclysmic Variable Nova-Likes}

Cataclysmic variables (CVs) are close binaries in which the primary, a
white dwarf (WD), accretes matter and angular momentum from the
secondary, a main sequence star, filling its Roche-lobe.  The matter
is transferred by means of either an accretion disk around the WD, or
an accretion column, when the WD has a strong ($\sim 10$ MG) magnetic
field.  Ongoing accretion at a low rate (quiescence) is interrupted
every few weeks to months by intense accretion (outburst) of days to
weeks (a dwarf nova accretion event), and every few thousand years by
a thermonuclear explosion (TNR; the classical nova event).  CV systems
are divided into sub-classes according to the duration, occurrence and
amplitude of their outburst: dwarf novae (DN) are found mostly in the
low state; while nova-likes (NL) exhibit the spectroscopic and
photometric characteristics of novae between or after outbursts but
have never had a recorded outburst.  NLs form a less homogeneous class
and are divided into sub-types depending on certain properties, such
eclipsing systems, magnetic systems, or systems that go into
unexpected low states \citep{war95}.  Non-magnetic NLs can be divided
into UX UMa systems which remain in a high optical brightness state or
"permanent outburst state", and the VY Scl systems (anti-dwarf novae)
which experience unexpected low states when the optical brightness
plummets.

CVs are believed to evolve from long period ($\sim $2 day) toward
short period ($\sim$1 hour) as the mass ratio $q$ decreases, crossing
the 2--3 hour 'period gap', where few system are found.  Theory
predicts that the period of those systems having a sufficiently low
$q$ will increase as $q$ continues to decrease, the so-called 'period
bouncers' \citep{pat05b}.  The driving mechanism behind the mass
transfer ($\dot{M}$) and hence, the evolution of CVs is believed to be
angular momentum loss, dominated by magnetic stellar wind braking at
periods above 3 hours and by gravitational wave emission below this
period.  The secondaries are stripped to $< 0.08 M_{\odot}$ on a
timescale of only 1-4 Gyr after they form as close binaries above the
gap.  Consequently, in the lifetime of the galaxy, the vast majority
of CVs should have evolved to a period minimum near 80 min and now
have degenerate brown dwarf-like secondaries \citep{how01}.  However,
recent models \citep{and03} suggest (i) much lower angular momentum
loss rates, so it takes 10-12 Gyr to even reach a 3 hr period, and
(ii) multiple evolutionary tracks yielding different populations of
CVs above and below the 3 hour orbital period.

The accretion disks and underlying accreting WDs in CVs can provide
crucial clues and constraints on these evolutionary scenarios. It is
important to derive $\dot{M}$ from observations, as the mass transfer
rate are tied to the angular momentum loss.  As to the WDs, they are
the central engines of the observed outbursts, either as potential
wells for the release of accretional energy or as the sites of
explosive TNR shell burning.  In addition, the differences between the
temperatures, rotation, and chemical abundances of CV WDs and single
isolated WDs provide clues as to the effects of accretion, diffusion,
and long term heating and evolution \citep{sio95}.

With a wavelength range from 905--1187\AA\AA, {\it FUSE} covers that
part of the far-ultraviolet (FUV) spectrum where the hotter CV
component is dominant.  For nova-likes, the dominant component in the
{\it FUSE} range is usually the accretion disk.  Additional possible
components are the hot white dwarf, the boundary layer or accretion
belt, and other hot regions on or close to the WD.  Many CV FUV
spectra are modeled with 2 components, usually a WD plus a disk.  For
each system, one usually derives from the {\it FUSE} spectrum the
temperature of the WD ($T_{eff}$), the gravity, the rotational
velocity ($V_{rot} \sin{i}$), the chemical abundances, the mass
accretion rate, the inclination, and the distance to the system.

It has therefore been very important to try and obtain {\it FUSE} data
for more CV systems in order to populate the $T_{eff}$-Period and the
$\dot{M}$-Period parameter spaces.  There is a need to enlarge the
sample to obtain an accurate global picture of the systems above and
below the period gap.  For that reason, we proposed to observe with
{\it FUSE} of a set of 16 high declination dwarf novae (our Cycle 7
survey of DNe) and 16 high declination nova-likes (our Cycle 8 survey
of NLs - all chosen from the on-line CV catalog of Downes {\it et
al.}). Unfortunately, our {\it FUSE} NL survey was cut short due to
the fatal failure of the reaction wheel of the telescope.  Of the 16
targets, only two were observed: BB Dor and P831-57.  The analysis of
the {\it FUSE} spectrum of P831-57 is presented elsewhere
\citep{bar08}.  In this work we present the result of the {\it FUSE}
spectroscopic analysis of BB Dor.

\subsection{The Nova-like BB Doradus} 

BB Dor (also known as EC 05287-5847, an object from the
Edingburgh-Cape Survey \citep{che01}) was spectroscopically identified
as a CV in December 1987 and was seen fainter than $V \sim 16.5$
until November 1992 \citep{che01}.  Since that time it seems to be in
bright state ($V\sim 14.6-13.6$).  A first (extremely uncertain)
estimate of its period by \citet{che01} put it right in the middle of
the period gap with a period of 0.107 day or 2.57 h.  However, a 45
day coverage \citep{pat02,pat05a} corrected this value to 0.14923
d or 3.58 h, as expected for VY Scl stars which usually have periods
between $\sim 3$ and $\sim 4$ h.  BB Dor exhibits superhumps with a
period excess of $\varepsilon=$9.4\% \citep{pat02} --- it has the largest 
known fractional superhump of all systems
and therefore provides a calibrating point 
at large $\varepsilon$ \citep{pat05a}.  
BB Dor is a VY Scl nova-like and so far no magnetic activity has been found.
The system also exhibits quasi-periodic oscillations \citep{che01}.
\citet{che01} also noted that due to its very narrow Balmer lines
(with wings extending $\sim 650$km$~$s$^{-1}$ from the line center) BB
Dor is most likely a low-inclination system.  From the Austral
Variable Star Observer Network (AVSON), it appears that BB Dor has
been around $V \approx 13.6-14.6$ mag (with an error $<0.1$ mag) for
the last couple of years and can vary as much as 0.4 mag within one
hour.  The reddening value E(B-V) towards BB Dor is not known.  
However, the Galactic reddening value in the direction 
of BB Dor as inferred from the 100$\micron$ dust emission map \citep{sch98} 
is very small: E(B-V)=$\sim$0.03.  Since the dust map
gives the maximum dust extinction, we deduce that BB Dor must have a
negligible reddening ($E(B-V) \sim 0.01$ at most) which should not 
affect the results; consequently we
do not deredden its {\it FUSE} spectrum.

\section{Observations}

\subsection{The {\it FUSE} Spectrum of BB Dor}

BB Doradus was observed with {\it FUSE} on 09 July 2007 at 02h51m09s
(UT), and 3 exposures were obtained (H9030301 - exposures 001, 002,
003). The {\it{FUSE}} spectrum was obtained through the low resolution
(LWRS; 30"x30") aperture in TIME TAG mode.  The data were processed
with the last and final version of CalFUSE v3.2.0 \citep{dix07}.

A look at the count rate graph (generated by CALFUSE) indicates that
the data for each exposure were collected mainly during the first 1000
s, followed by a jitter lasting about 1,500 s.  Toward the end of each
exposure, the count rate dropped to zero.  Therefore, while the
co-added exposures for the 8 individual segments after processing by
CalFUSE give a good exposure time ({\it after screening}) of between
about 8750 s (segments 1) and 9750 s (segments 2), the count rate plot
indicates that the total {\it good collection time} ranges between
3350 s (segments 1) and 3650 s (segments 2).  We therefore weighted
these segments accordingly, multiplying the flux (of the order of $1
\times 10^{-13}$ergs$~$s$^{-1}$cm$^{-2}$\AA$^{-1}$ in the on-line {\it
preview}) by a factor of 2.61 (segments 1) and 2.67 (segments 2).
After that we processed the data as usual (see below).

The spectral regions covered by the spectral channels overlap, and
these overlap regions are then used to scale the spectra in the SiC1,
SiC2, and LiF2 channels to the flux in the LiF1 channel.  The low
sensitivity portions (usually the edges) of each channel are
discarded.  In the present case the SiC2b spectral segment was very
noisy and we discarded it too.  We also carried out a visual
inspection of the {\it{FUSE}} channels to locate the worm and we
{\it{manually}} discarded these portions of the spectrum affected by
the worm.  We combine the individual exposures and channels to create
a time-averaged spectrum weighting the flux in each output datum by
the exposure time and sensitivity of the input exposure and channel of
origin.  The final product is a spectrum that covers the full
{\it{FUSE}} wavelength range $905--1187$\AA\AA.  Because we
disregarded the right edge of both the SiC1a and SiC2b segments, there
is a gap between the LiF1a and LiF2b segments ($\sim 1082$ -- $\sim
1087$\AA\AA).

The {\it FUSE} spectrum of BB Dor is shown in Figure 1.  Note
that the {\it FUSE} exposures of BB Dor
consist entirely of observations carried out during the DAY. The small
NIGHT exposure was ignored because of jitters and other problems.  We
suspect that most of the emission lines are due to helio- and
geo-coronal emissions.

\subsection{The {\it{FUSE}} Lines} 

The {\it{FUSE}} spectra of CVs exhibit mainly broad absorption lines
from the accretion disk and WD, as well as sharp absorption lines from
circumstellar (or circumbinary) material and the ISM.  Emission lines
from the hotter regions are usually broadened, depending on the
inclination angle $i$, due to the large Keplerian velocity in the
inner disk.  Sharp emission lines from air glow (geo- and
helio-coronal in origin) are also present, due mainly to the
reverberation of sunlight inside the {\it FUSE} telescope during DAY
time observations. The present {\it FUSE} spectrum of BB Dor presents
such a complexity, and all the lines are listed in Table 1.

\paragraph{Absorption Lines.} 
The main characteristic of the {\it{FUSE}} spectrum of BB Dor is the
broad Ly$\beta$ absorption feature due to either the exposed WD, the
disk at low inclination angle $i$, or possibly both (see the Results
section).  Absorption features from the higher orders of the Lyman
series are also clearly visible, indicating a temperature
T$\sim35,000$K.  The other main absorption features expected at this
temperature are the C\,{\sc iii} (1175 \AA), Si\,{\sc iii}
($\approx$1108--1114\AA\AA\ and $\approx$1140--1144\AA\AA), Si\,{\sc
iv} ($\approx$1120--1130\AA\AA) and the C\,{\sc ii} (1010\AA).  The
spectrum, is however, also marked by some high-order ionization
absorption lines such as S\,{\sc iv} (1063 \& 1073\AA), Si\,{\sc iv}
(1066.6\AA), and the O\,{\sc vi} doublet.  The S\,{\sc iv} (1006\AA),
C\,{\sc ii} (1010\AA), and S\,{\sc iv} (1099\AA) absorption multiplets
are all unresolved and shallow.  However, we have indicated them on
the figures, as they appear deeper in the modeling.  On the other side
the S\,{\sc iv} (1063\AA\ \& 1073\AA) and Si\,{\sc iii}
($\sim1010$\AA) absorption lines appear deeper in the observed
spectrum than in the modeling.  The presence of the high-order
ionization species (such as the oxygen doublet) indicates the presence
of a hotter absorbing component above the main FUV emitting region.
The P\,{\sc ii} line ($\sim$961\AA ) has been marked in the spectrum,
though it is most probably a feature due to the noise. 
All the the lines are listed in Table 1 with their wavelengths.
Most of the broad absorption lines and features associated with
the source are redshifted by about 0.6-0.9\AA , corresponding to
a receding velocity of $200 \pm 40$km$~$s$^{-1}$. It seems very
likely that this redshif is due to the orbital motion of the WD,
however, we do not have enough time tag data to verify this assumption.

\paragraph{Emission Lines.} 
The spectrum of BB Dor exhibits sharp emission lines including all the
orders of the Lyman series, N\,{\sc iv} ($\sim$923\AA), S\,{\sc vi}
(933.5\AA\ \& 944.5\AA), C\,{\sc iii} (977\AA), He\,{\sc ii} (992\AA),
and He\,{\sc i} (1068\AA).  
Some of these lines seems to be due to
helio-coronal emission (sunlight reflected inside the telescope) that
contaminates the SiC channels, in particular the H\,{\sc i} lines, the
He\,{\sc i} \& He\,{\sc ii} lines, the C\,{\sc iii} (977\AA) line, and
the O\,{\sc vi} doublet.  
The segments of the spectrum from the LiF
channels overlapping the SiC Channels down to about 1000\AA\ do not
show any oxygen lines, and we therefore did not include these spectral
regions of the SiC channels.  However, other emission lines may be
solely from the source. In many NLs, the emission lines from the disk
are broadened by the Keplerian velocity and are consequently easily
identified.  However, \citet{che01} have detected very narrow Balmer
lines and concluded that BB Dor is most likely a low-inclination
system.  Because of this, the identification of the emission lines
from the source is not trivial.  This issue is further complicated by
the lack of NIGHT exposure and the very low signal-to-noise ratio,
which makes it difficult to analyze the {\it two-dimensional image} of
the spectrum for the effect of scattered sunlight on the emission
lines.

Since the system is a NL in a high state, we expect at first the N\,{\sc iv}
($\approx$923\AA), S\,{\sc vi} (933.5 \& 944.5 \AA), and C\,{\sc iii}
(977\AA) emission lines to be from the source with some possible
helio-coronal contamination mostly affecting the carbon emission
lines.  The N\,{\sc iv} (923.06\AA) emission line is contaminated with
the H\,{\sc i} (923.15\AA) line; the N\,{\sc iv} (922.52\AA\ +
924.28\AA) lines are not.  
However, the complete absence of narrow emission lines from the 
oxygen doublet and the C\,{\sc iii} (1175) seems to indicate 
otherwise --- i.e. all the sharp emission lines might be helio-coronal
in origin. To further check this possibility we measured the narrow
emission line relative intensities below 1000\AA and compared 
with those of the solar spectrum \citep{cur01}. We 
found   a good agreement with the solar-disk average quiet Sun data,  
except for
the SVI 944.5 which has a higher relative intensity in the spectrum
of BB Dor. This could be due sun spot activity during the FUSE observation
of BB Dor, as this line intensity increases by a factor of 30 inside
sun spots while the other lines increase at most by a factor 3.
We therefore conclude that all the sharp emission lines in the {\it FUSE} 
spectrum of BB Dor are due to sun light reflected inside the telescope.

\section{Spectral Modeling} 

We created a grid of models of synthetic spectra of WDs and accretion
disks for different values of the WD temperature $T_{eff}$, gravity
$Log(g)$, projected rotational velocity $V_{rot} \sin(i)$, inclination
$i$, mass accretion rate $\dot{M}$, and abundances.  We then ran a
$\chi^2$ fitting program to find the best fit for (i) a single WD
component, (ii) a single accretion disk component, and (iii) a
combined WD + accretion disk.  We describe how we generate these
synthetic spectra and how we perform the fitting.

We create the synthetic model spectra for high-gravity stellar
atmospheres using codes TLUSTY and SYNSPEC\footnote{
http://nova.astro.umd.edu; TLUSTY version 200, SYNSPEC version 48}
\citep{hub88,hub95}.  Atmospheric structure is computed (using TLUSTY)
assuming a H-He LTE atmosphere; the other species are then added in
the spectrum synthesis stage using SYNSPEC.  We generate photospheric
models with effective temperatures ranging from $\sim$20,000--50,000 K
in increments of 1,000 K.  We chose values of $Log(g)$ ranging between
7.5 and 9.0.  We also varied the stellar rotational velocity $V_{rot}
sin(i)$ from $100$km$~$s$^{-1}$ to $1000$km$~$s$^{-1}$ in steps of
$100$km$~$s$^{-1}$ (or smaller if needed).  The WD rotation ($V_{rot}
sin(i)$) rate is determined by fitting the WD model to the spectrum
while paying careful attention to the line profiles in the {\it{FUSE}}
spectrum.  We do not carry out separate fits to individual lines but
rather try to fit the lines and continuum simultaneously while paying
careful attention to the absorption lines.  It is important to note
that the depth of the absorption features depends not only on the
abundances but also on the rotational velocity.  Increasing the
rotational velocity reduces the central depths of the absoption 
features, thus reducing the abundance. However, the widths of the
absorption features also increase with increasing rotational velocity.  
As a consequence, abundances and rotational velocity effects
are interwined and cannot always be easily separated without
additional information on either parameter. 
For any WD mass there is a corresponding radius, or
equivalently one single value of $Log(g)$ (e.g. see the mass radius
relation from \citet{ham61}, or see \citet{woo90,pan00} for different
composition and non-zero temperature WDs).  Therefore, by scaling the
theoretical spectrum to the observed one, we obtain the distance to
the system.

We model accretion disk spectra by first assuming that the disk is
made of a collection of annuli, where each annulus has a temperature
$T(r)$ and gravity $Log(g(r))$ given by the standard disk model
\citep{sha73,pri81}, for a given central mass $M_{WD}$ and accretion
rate $\dot{M}$.  A variant of the code TLUSTY is then used, TLDISK,
that generates an atmosphere model for each annuli that is then used
as input for the code SYNSPEC.  The contribution of all the annuli are
then combined using the code DISKSYN, and a final spectrum is obtained
for any given inclination angle.  A detailed explanation of the
procedure is given in \citet{wad98}.  In the present work we do not
use the grid of synthetic accretion disk spectra tabulated by
\citet{wad98}, instead we generate them.  This allows us to compute
disk spectra assuming non-solar abundances and for any inclination
angle (the disk spectra of \citet{wad98} have solar abundances and
have been generated for given value of the inclination $i$).

Before carrying out a synthetic spectral fit of the spectra, we masked
portions of the spectra with strong emission lines, strong ISM
absorption lines, detector noise, and air glow.  The regions excluded
from the fit are in blue in Figures 1 and 2.

After having generated grids of models for the {\it FUSE} spectrum of
BB Dor, we use FIT \citep{numrec}, a $\chi^2$ minimization routine, to
compute the reduced $\chi^{2}_{\nu}$ ($\chi^2$ per number of degrees
of freedom) and scale factor (which gives the distance) for each model
fit.  While we use a $\chi^2$ minimization technique, we do not
blindly select the least $\chi^2$ models, but we also examine the
models that best fit some of the features such as absorption lines
and, when possible, the slope of the wings of the broad Lyman
absorption features.

Initially we generate solar abundances models, and when a good fit is
found, we start varying the chemical abundances of C, N, S and Si, to
fit the absorption features of the spectrum.  In particular, the
carbon abundance was set using the C\,{\sc ii} (1010\AA) and C\,{\sc
iii} (1175\AA) multiplets; the sulfur abundance was set using the
S\,{\sc iv} (1063\AA\ \& 1073\AA) lines; and the silicon was set using
the Si\,{\sc iv} (1067\AA\ , 1023\AA\ , 1028\AA) and Si\,{\sc iii}
($\sim$1110\AA\ , $\approx$1138-1146\AA) lines.

\section{Results}

The data obtained by the AVSON implies that BB Dor has been in a high
state around $V \approx 13.6-14.6$ for the last couple of years.  We
therefore expect the {\it FUSE} spectrum to be dominated by flux from
the accretion disk with a relatively high $\dot{M}$.  However, in our
modeling we follow a systematic approach which consists in fitting the
following: (i) a single WD; (ii) a single accretion disk; and (iii) a
WD + accretion disk composite.

\subsection{White Dwarf.} 
Since we do not have any information on the mass of the WD, we look
for all the best-fit models in the $Log(g)$ versus $T_{eff}$ plane.
Namely, for each assumed value of $Log(g)$ we vary the temperature to
find the best-fit model.  As expected we find that the temperature is
somewhere between 34,000K (for $Log(g)=7.5$) and 40,000K (for
$Log(g)=9.0$), with a distance between 377 pc and 135 pc, respectively
(see Table 1).  The least $\chi^2_{\nu}$ is obtained for
$Log(g)=8.0-8.65$.  We chose the intermediate value $Log(g)=8.3$
model, with T=37,000K and d=247 pc and solar abundances, to illustrate
our results in Figure 1.  We then further improve the fit by varying
the abundances.  We find that in order to better fit the sulfur and
silicon lines, we have to set the sulfur to 20 times its solar
abundance, and the silicon to 3 times, while all the other species are
kept at solar abundances.  The $\chi^2_{\nu}$ value decreases from
0.3348 to 0.3129 (by about 7\%).  While some of the lines are better
fitted (such as S\,{\sc iv} 1063\AA\ \& 1073\AA; Si\,{\sc iii}
$\sim$1010\AA\ \& Si\,{\sc iv} $\sim$1025\AA), the S\,{\sc iv}
(1006\AA\ \& 1100\AA) are far too deep (see Figure 2).

\subsection{Accretion Disk.} 
Next, we fit the solar abundances disk models.  We find that a low
inclination is needed in order to match the absorption lines, and in
our models we initially set $i$=5, 8, 12, and 18 degrees.  Again,
because we have no information on the WD mass, for each value of
$Log(g)$ (ranging between 7.5 and 9.0) we vary the mass accretion rate
between $10^{-10.5}M_{\odot}$yr$^{-1}$ and
$10^{-8.0}M_{\odot}$yr$^{-1}$ to find the best-fit models.  Again we
find that the best-fit models are around $Log(g) = 8.3$, but the
difference between the $\chi^2_{\nu}$ values is not significant, and
the improvement over the single WD models is also only marginal (of
the order of $\sim$1\% in the $\chi^2_{\nu}$ value).  All these models
are presented in Table 2.  Next, we decide to improve the disk model
by varying the abundances, though we do not expect the fit to improve
much because of the Keplerian velocity broadening of the disk.
However, we find that setting the silicon and sulfur to 3 and 20 times
solar, resp., actually reduced the $\chi^2_{\nu}$ value by 10\%, more
than for the single WD model.  This is because the the S\,{\sc iv}
(1006\AA\ \& 1100\AA) lines have better fits (Figure 3).  This
best-fit single disk model has $i=8^{\circ}$, $\dot{M} = 10^{-9}
M_{\odot}$yr$^{-1}$, a distance of 665 pc and $\chi^2_{\nu}=0.3000$.

\subsection{Composite Model: WD + Accretion Disk.} 

Last, we fit composite WD+Disk models.  Since the number of models
increases exponentially when we change $Log(g)$, $T_{eff}$ and
$\dot{M}$, we restrict our search using our best fit for $Log(g)$,
namely 8.3.  We also set the sulfur to 20 times solar and silicon 3
times solar.  This procedure reduces the number of free parameters to
3 ($\dot{M},T_{eff},i$).

\paragraph{Low Inclination.} Since the
system is believed to have a low inclination angle \citep{che01}, we
generate low-inclination models ($i=5^{\circ},~8^{\circ},~
12^{\circ},$ \& $18^{\circ}$).  Not surprisingly, the best fit model
has $i=8^{\circ}$ and $\dot{M}=10^{-9}M_{\odot}$yr$^{-1}$, but this
time the WD has a temperature of 32,000K and provides only 10\% of the
flux, while the disk provides the remaining 90\%.  Such a model again
brings an insignificant improvement in the value of $\chi^2_{\nu}$ and
it is clear (Figure 4) that it is barely distinguishable from the
best-fit single disk model.  From the point of view of the physics,
this model is preferred because, if the system has a low-inclination
angle, then the emission from the WD must contribute to the flux.
Models with a WD temperature $T<30,000$K had a WD contribution of only
a few percent of the total FUV flux, and could not be distinguished
from the single disk models.  If the WD has a temperature
$T_{eff}<$30,000K it will not be detected while the system is in a
high state (with $\dot{M}=10^{-9}M_{\odot}$yr$^{-1}$).  These results
implies that the contribution from the WD is not very large and that
the temperature of the WD must be $\leq 32,000$ K.

\paragraph{High Inclination.} 
From the sharp Balmer emission lines, \citet{che01} suggest the system
has a low inclination.  However, we cannot confirm the inclination
directly from the emission lines of the {\it FUSE} spectrum, as {\it
all} the sharp emission lines in the {\it FUSE} spectrum are of
helio-coronal origin.  Our low-inclination angle single disk models
provide a slightly better fit than the single WD models (and a much
better fit than the high-inclination single disk models), however, for
completeness we include here the results from the composite WD + disk
model fits when the assumption on the inclination angle is relaxed.

We search for the best-fit WD+disk models (assuming $Log(g)=8.3$) in
the parameter space $T_{eff}$ versus $\dot{M}$ including {\it all}
inclination angles, and find that the models with an intermediate
inclination ($i=18^{\circ}$, $41^{\circ}$, $60^{\circ}$, \&
$75^{\circ}$) do not provide the best fit.  The best-fit model has
$i=80^{\circ}$, and reflects a situation in which the WD is dominant
with $T=37,000$K and contributes 3/4 of the total flux, while the disk
has $\dot{M}=10^{-9}M_{\odot}$yr$^{-1}$ and contributes only 1/4 of
the flux.  The distance obtained from this model is 246 pc and
$\chi^2_{\nu}=0.2873$.  This is the least $\chi^2$ value we obtained
from all our models.  This model is presented in figure 5.  Similar
results were obtained when assuming $Log(g)=8.65$, but with a slightly
lower mass accretion rate (Table 2).

\section{Discussion and Conclusion} 

BB Dor is a little-known southern NL, and, consequently, both the
distance and the mass of the WD are unknown, which implies a larger
uncertainty in the results.  In addition the {\it FUSE} spectrum is
definitely of poor quality.  In theory, a fine tuning of the
temperature (say to an accuracy of about $\pm$50 K) and mass accretion
rate can be carried out by fitting the flux levels such that the
distance to the system (when known) is obtained accurately.  However,
the fitting to the distance depends strongly on the radius (and
therefore the mass) of the WD.  We discuss some additional
restrictions that we use to constrain the properties of the system.

On the basis of the least $\chi^2_{\nu}$, the best model is the high
inclination WD+disk composite.  However, (see Figure 5), because of
the high inclination, the disk contributes a rather flat component all
the way into the shorter wavelengths. At such a high inclination, one
would not expect the WD to dominate the flux, but rather the WD would
be almost completely masked by the swollen disk; actually the
system would likely be observed to undergo eclipses --- but none
have been observed. Also, since BB Dor
can vary as much 0.4 mag in one hr, this means that the light cannot
be dominated by the WD, and the disk must be contributing at least
40\% of the light. In other words, while the least $\chi^2$ indicates
that the best fit is a WD with a rather flat disk component, there are
other indications that this cannot be correct.  Actually, the WD+disk
high inclination model strikingly resembles the second component
observed in the {\it FUSE} spectra of some dwarf novae during
quiescence (e.g. VW Hyi, \citet{god04}).  This is likely a caveat in
the {\it state-of-the-art} spectral modeling, rather than an
indication of a physical link between the the spectrum of BB Dor and
that of a DN in quiescence.  The need for improved modeling also stems
from the difficulty in producing a model that fits low and high-order
ionization lines at the same time.  For BB Dor it is possible that the
S\,{\sc iv} 1063\AA\ \& 1073\AA\ absorption lines form in the same
hotter region/layer where the O\,{\sc vi} doublet form; while the
C\,{\sc iii} and Si\,{\sc iii} lines form in a cooler region/layer
where the Lyman series (and continuum) form.  This is similar to the
HST/STIS spectrum of TT Crt \citep{sio08} which exhibits a rich
variety of absorption lines from different ionization stages,
suggesting line formation in (at least) two different temperature
regions.

In order to reduce the size of the domain for which we have best-fit
models in the parameter space, we use the infrared magnitudes J, H, \&
K from the {\it Two Micron All Sky Survey} (2MASS) to assess the
distance to BB Dor as prescribed by \citet{kni06} for systems with
$P<6$ h.  The IR data were collected on 9 November 1999, at a time
when BB Dor was in a high state with a visual red magnitude R=14.60
and a blue magnitude B=13.90 (whereas B$\sim$16.5 in the low state).
The {\it 2MASS} IR apparent magnitudes are J=14.322, H=14.089, and
K=14.053.  For a primary star $M_{wd}=0.75 M_{\odot}$ and period of
3.559 h (corresponding to BB Dor), the donor star mass is 
$M_2=0.25 M_{\odot}$ 
(\citet{pat05a} suggest $M_2=0.256 M_{\odot}$)  
and the IR absolute magnitude estimates are $M_J=7.47$,
$M_H=6.90$ and $M_K=6.63$.  Inserting these into eq.(15) of
\citet{kni06} gives a distance of 235, 274 \& 305 pc, respectively.
These distances are typically underestimated by factors of 2.05, 1.86
and 1.75 for the J, H and \& K bands giving distances of 482, 510, and
534 pc, respectively, assuming the donor star contributes only $\sim
1/4 - 1/3$ of the total IR flux \citep{kni06}.  The distance to BB Dor
is therefore certainly larger than 300 pc, and most likely in the
range of $\sim 500$ pc.  Since BB Dor was observed in a high state
(with 2MASS), it is likely that the infrared flux of the donor star is
on the lower side, namely even less than 1/4, in which case the 500 pc
itself is only a lower limit.  Using this restriction on the distance,
we see that we can safely reject the single WD models (with
$d<400$pc), the high inclination models (with $d<250$pc), and the
large mass $Log(g)=9.0$ models (with $d=361$pc and less).

On the basis of the distance alone, the low inclination disk and
WD+disk models are the best fits.  These are also the best fit on the
basis of the optical spectrum (low inclination) and visual magnitude
(high state/disk).

We are now able to summarize the basic characteristics of the system
and its WD.  We are confident in the fact that the system must have a
WD with $Log(g) = 8.3\pm0.3$, $T_{eff}=32,000$K {\it or lower}
(disk+WD composite models), a mass accretion rate of the order of
$\dot{M}=10^{-9}M_{\odot}$yr$^{-1}$ (disk and disk+WD models) and a
distance somewhere between 500 and 700 pc (choosing the least $\chi^2$
models agreeing with the distance).  From the best-fit single disk
models alone we find that the inclination angle must be very small
($\sim8^{\circ}$) to fit the absorption lines.  There is an indication
that the sulfur and silicon are over abundant and it is also possible
that some of the high-order ionization lines form in a layer in front
of the disk and the WD.

Although the {\it FUSE} spectrum of BB Dor is rather poor and
multiwavelength data are limited, we have narrowed down the region in
the parameter space ($\dot{M}$, $P$, $T_{eff}$) for this system. BB
Dor is now the sixth VY Scl NL variable with a temperature estimate
for its WD. With a temperature of $\leq 32,000$ K, the WD of BB Dor
marks the lower end of the temperature distribution for WDs of VY Scl
NL variables, which, so far, for the 5 other systems, ranged between
40,000K and 47,000K \citep{ham08}.  The {\it FUSE} spectrum of BB Dor
clearly shows the drop in flux in the shorter wavelengths, the
signature of a moderate temperature when compared to the {\it FUSE}
spectra of these other VY Scl systems (see e.g. V794 Aql,
\citep{god07}).  Even if we consider the single WD best-fit model, the
WD temperature of BB Dor only reaches 37,000K (and the 40,000K
$Log(g)=9.0$ model gives an unrealistic close distance of 135pc in
disagreement with the expected IR emission from the secondary).  In
that respect, the data point in the ($T_{eff},P$) plane for BB Dor
falls {\it under} the region of the VY Scl NL variables, seemingly
{\it into} the region of the parameter space populated with dwarf
novae \citep{sio08}.  However, as shown in Figure 6, the data point
for BB Dor does not seem to stand apart, and the separation between VY
Scl systems and DN can be made easily by drawing a diagonal line. In
that respect, it is actually V794 Aql that divides the two regions of
the graph as indicated by the slanted line.

\acknowledgments
We thank the anonymous referee for a prompt, concise
and pertinent report which helped improve the analysis of the
lines.
Special thanks to the Austral Variable Star Observer Network (AVSON),
and in particular to Berto Monard for providing us with the visual
magnitude of BB Dor between 02 Sep., 2005 and 30 Dec., 2007, and to
Peter Nelson for a quick look V band photometry at BB Dor on 28 Jan.,
2008.  PG wishes to thank Mario Livio, for his kind hospitality at the
Space Telescope Science Institute, where part of this work was done.
The infra-red J, H, \& K magnitudes of BB Dor were retrieved from the
online archival data of the {\it 2MASS} at IPAC.  
Part of this publication makes use of the data products from the Two Micron
All Sky Survey, which is a joint project of the University of
Massachusetts and the Infrared Processing and Analysis Center/California
Institute of Technology, funded by the National Aeronautics and
Space Administration and the National Science Foundation.
This research was
based on observations made with the NASA-CNES-CSA Far Ultraviolet
Spectroscopic Explorer.  {\it{FUSE}} is operated for NASA by the Johns
Hopkins University under NASA contract NAS5-32985. This research was
supported by NASA {\it FUSE} grant NNX07AU50G to Villanova University
(P. Godon).  This work was also supported in part by NSF grant
AST05-07514 to Villanova (E.M. Sion).

\begin{deluxetable}{cccc} 
\tablewidth{0pc}
\tablecaption{ {\it FUSE} Line Identification} 
\tablehead{
Ion          & $\lambda_{rest}$ (\AA) & $\lambda_{obs}$ (\AA ) & Comments$^a$ } 
\startdata 
H\,{\sc i}   &    918.13 &  918.0    &  c/e   \\
             &    919.35 &  919.3    &  c/e   \\
             &    920.96 &  920.9    &  c/e   \\
N\,{\sc iv}  &    922.52 &  922.3    &  s/e    \\
             &    923.06 &  923.0    & s,c/e   \\
H\,{\sc i}   &    923.15 &  923.2    &  c/e   \\
N\,{\sc iv}  &    923.22 &  923.2    & c/e   \\
             &    924.28 &  924.2    & c/e   \\
H\,{\sc i}   &    926.23 &  926.2    &  c/e   \\
             &    930.75 &  930.7    &  c/e   \\
S\,{\sc vi}  &    933.50 &  933.4    &  c/e     \\
H\,{\sc i}   &    937.80 &  937.8    &  c/e    \\
S\,{\sc vi}  &    944.50 &  944.4    &  c/e    \\
H\,{\sc i}   &    949.74 &  949.7    &  c/e     \\
N\,{\sc i}   &    952.40 &  925.5    & ism/a     \\
             &    953.42 &  ---      &  ---      \\
             &    953.66 & ---       &  ---      \\
             &    953.97 & ---       &  ---      \\
             &    954.10 & 954.2     & ism/a   \\
P\,{\sc ii}  &    961.04 & 961.2     & ism/a?   \\  
H\,{\sc i}   &    972.54 &  972.5    &  c/e    \\
C\,{\sc iii} &    977.02 &  977.0    & c/e    \\
Si\,{\sc iii}&    993.52 & 993.2     & s?/a \\ 
             &    994.19 & 994.9     & s?/a \\ 
             &    997.39 & 997.1     & s?/a \\ 
S\,{\sc iv}  &   1006.07 &  ---      & s/a - unresolved \\ 
             &   1006.39 &  ---      & s/a - unresolved \\ 
C\,{\sc ii}  &   1009.85 &  ---      & s/a - unresolved \\ 
             &   1010.08 &   ---     & s/a - unresolved \\ 
             &   1010.37 &   ---     & s/a - unresolved \\ 
H\,{\sc i}   &   1025.77 & 1025.7    &  c/e    \\  
O\,{\sc vi}  &   1031.91 & 1032.9    &  s/a     \\
C\,{\sc ii}  &   1036.34 & 1036.4    &  ism/a    \\
O\,{\sc vi}  &   1037.61 & 1038.3    &  s/a    \\
O\,{\sc i}   &   1039.10 & 1039.3    & ism/a    \\
Ar\,{\sc i}  &   1048.20 & 1048.3    & ism/a    \\
S\,{\sc iv}  &   1062.65 & 1063.4    &  s/a     \\
Si\,{\sc iv} &   1066.60 & 1067.1    &  s/a     \\
Ar\,{\sc i}  &   1066.66 & 1066.8    & ism/a     \\
S\,{\sc iv}  &   1072.97 & 1073.9    &  s/a   \\
             &   1073.52 & 1073.9    &  s/a \\
S\,{\sc iv}  &   1098.36 &  ---      & s/a - unresolved \\ 
             &   1098.93 &  ---      & s/a - unresolved \\ 
             &   1099.48 &  ---      & s/a - unresolved \\ 
             &   1100.53 &  ---      & s/a - unresolved \\ 
Si\,{\sc iii} &  1108.36 & 1108.8    &  s/a     \\
              &  1109.94 & 1110.6    &  s/a     \\
       &   $\sim$1113.20 & 1113.8    &  s/a     \\
Si\,{\sc iv}  &  1122.49 & 1123.0    &  s/a    \\
        &  $\sim$1128.33 & 1128.9    &  s/a    \\
N\,{\sc i}    &  1134.16 & 1134.3    & c/a  \\
              &  1134.42 & 1134.3    & c/a  \\ 
              &  1134.98 & 1135.2    & c/a    \\ 
S\,{\sc i}    &  1145.10 & 1145.0    & ism/a \\ 
He\,{\sc ii}  &  1168.61 & 1168.6    & c/e    \\
C\,{\sc iii}  &  1174.90 & 1176.4    & s/a - unresolved    \\
              &  1175.26 & 1176.4    & s/a - unresolved   \\  
              &  1175.60 & 1176.4    & s/a - unresolved    \\  
              &  1175.71 & 1176.4    & s/a - unresolved    \\  
              &  1176.00 & 1176.4    & s/a - unresolved    \\  
              &  1176.40 & 1176.4    & s/a - unresolved    \\  
\enddata
\tablenotetext{a}{The following abbreviations have been used: 
a - for absorption; e - for emission; c - for contamination 
(e.g. air glow; helio-coronal emission); s - for source; and
ism - for interstellar medium.}
\end{deluxetable} 

\begin{deluxetable}{cccccccccccc} 
\tablewidth{0pc}
\tablecaption{Synthetic Spectra} 
\tablehead{ 
$Log(g)$&$T_{wd}$&$V_{rot}sin{i}$&[Si] &[S] &[Z] & $i$   & $Log(\dot{M})$  &WD/Disk& d &$\chi^2_{\nu}$& Fig. \\ 
 (cgs)  &($10^3$K)&(km/s)&Solar & Solar & Solar  &  (deg) & ($M_{\odot}$/yr) &(\%)   &(pc)&              &      \\ 
} 
\startdata 
 7.50 & 34.0 & 400       & 1.0   & 1.0    & 1.0  &    ---   & ---     & 100/0 & 377      & 0.3427      &      \\
 8.00 & 36.0 & 400       & 1.0   & 1.0    & 1.0  &    ---   & ---     & 100/0 & 266      & 0.3361      &      \\
 8.30 & 37.0 & 400       & 1.0   & 1.0    & 1.0  &    ---   & ---     & 100/0 & 247      & 0.3348      &  1   \\
 8.30 & 37.0 & 400       & 3.0   & 20.   & 1.0   &    ---   & ---     & 100/0 & 211      & 0.3129      &  2   \\
 8.50 & 38.0 & 400       & 1.0   & 1.0    & 1.0  &    ---   & ---     & 100/0 & 193      & 0.3345      &      \\
 8.50 & 38.0 & 400       & 3.0   & 20.    & 1.0  &    ---   & ---     & 100/0 & 184      & 0.3100      &      \\
 8.65 & 38.0 & 400       & 1.0   & 1.0   & 1.0   &    ---   & ---     & 100/0 & 172      & 0.3360      &      \\
 9.00 & 40.0 & 400       & 1.0   & 1.0    & 1.0  &    ---   & ---     & 100/0 & 135      & 0.3390      &      \\
 7.50 & ---  & ---       & 1.0   & 1.0    & 1.0  &    12    & -8.0    & 0/100 & 878      & 0.3346      &      \\
 7.88 & ---  & ---       & 1.0   & 1.0    & 1.0  &    12    & -8.5    & 0/100 & 878      & 0.3346      &      \\
 8.30 & ---  & ---       & 1.0   & 1.0    & 1.0  &    8     & -9.0    & 0/100 & 693      & 0.3297      &      \\
 8.30 & ---  & ---       & 3.0   & 20.    & 1.0  &    8     & -9.0    & 0/100 & 665      & 0.3000      &  3   \\
 8.65 & ---  & ---       & 1.0   & 1.0    & 1.0  &    5     & -9.5    & 0/100 & 496      & 0.3332      &      \\
 9.00 & ---  & ---       & 1.0   & 1.0    & 1.0  &    5     & -10.0   & 0/100 & 361      & 0.3376      &      \\
 8.30 & 32.0 & 400       & 3.0   & 20.   & 1.0   &    08    & -9.0    & 10/90 & 700      & 0.2989      &  4   \\
 8.30 & 37.0 & 400       & 3.0   & 20.   & 1.0   &    80    & -9.0    & 74/26 & 246      & 0.2873      &  5   \\
 8.65 & 38.0 & 400       & 3.0   & 20.   & 1.0   &    80    & -9.5    & 75/25 & 184      & 0.2880      &      \\
\enddata
\end{deluxetable}

\clearpage 

\begin{figure}
\plotone{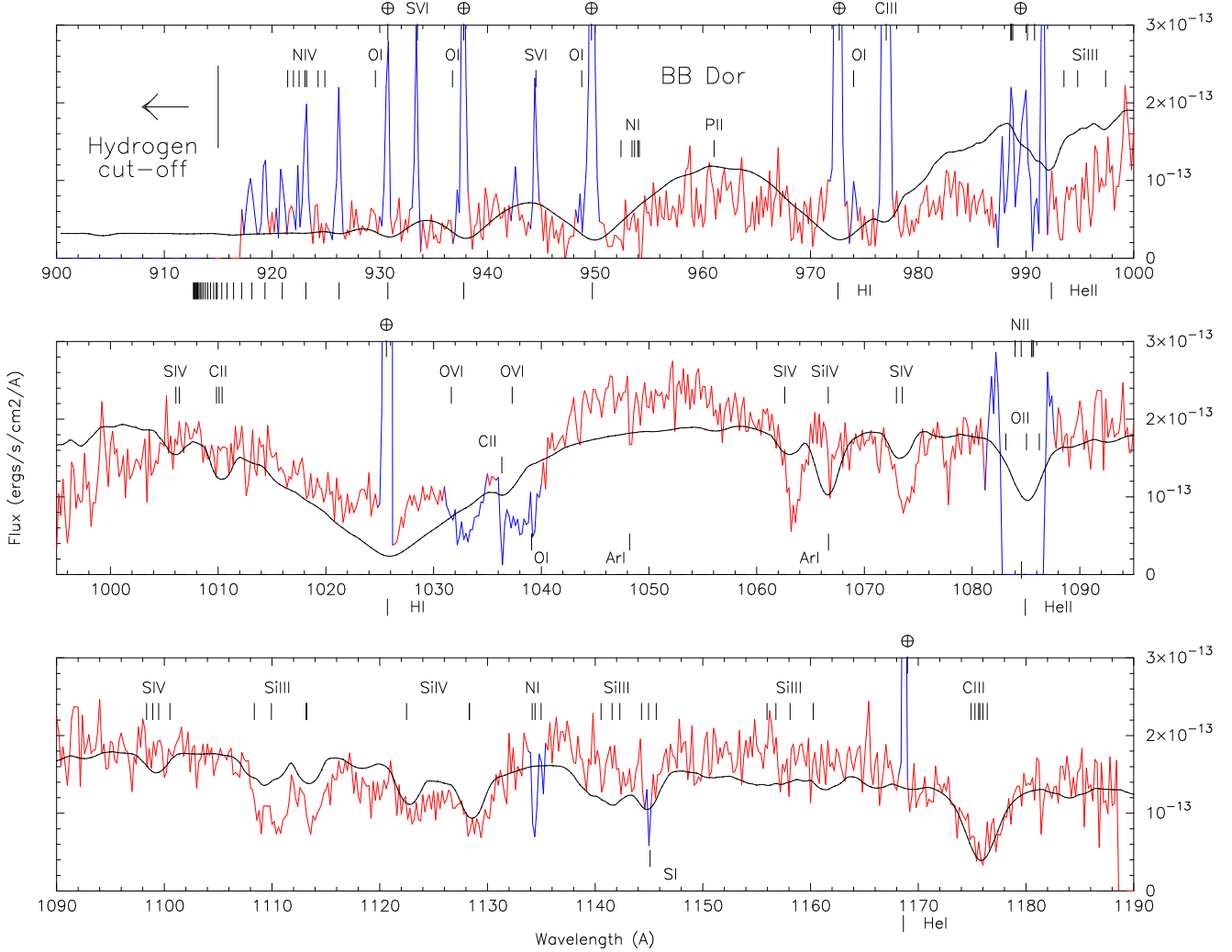}             
\vspace{-5.cm} 
\figcaption{Solar abundance single WD model.  The best-fit WD model
assuming solar abundances (solid black line) is shown together with
the {\it{FUSE}} spectrum of BB Dor (light grey/red line).  The
segments of the spectrum that have been masked before the fitting are
in dark grey/blue.  The WD has a temperature $T=37,000$K, a projected
rotational velocity $V_{rot} \sin{i} =400$km/s, and $Log(g)=8.3$. The
distance is $d=$217 pc, and $\chi^2_{\nu}=0.3348$.}
\end{figure} 

\clearpage 

\begin{figure}
\plotone{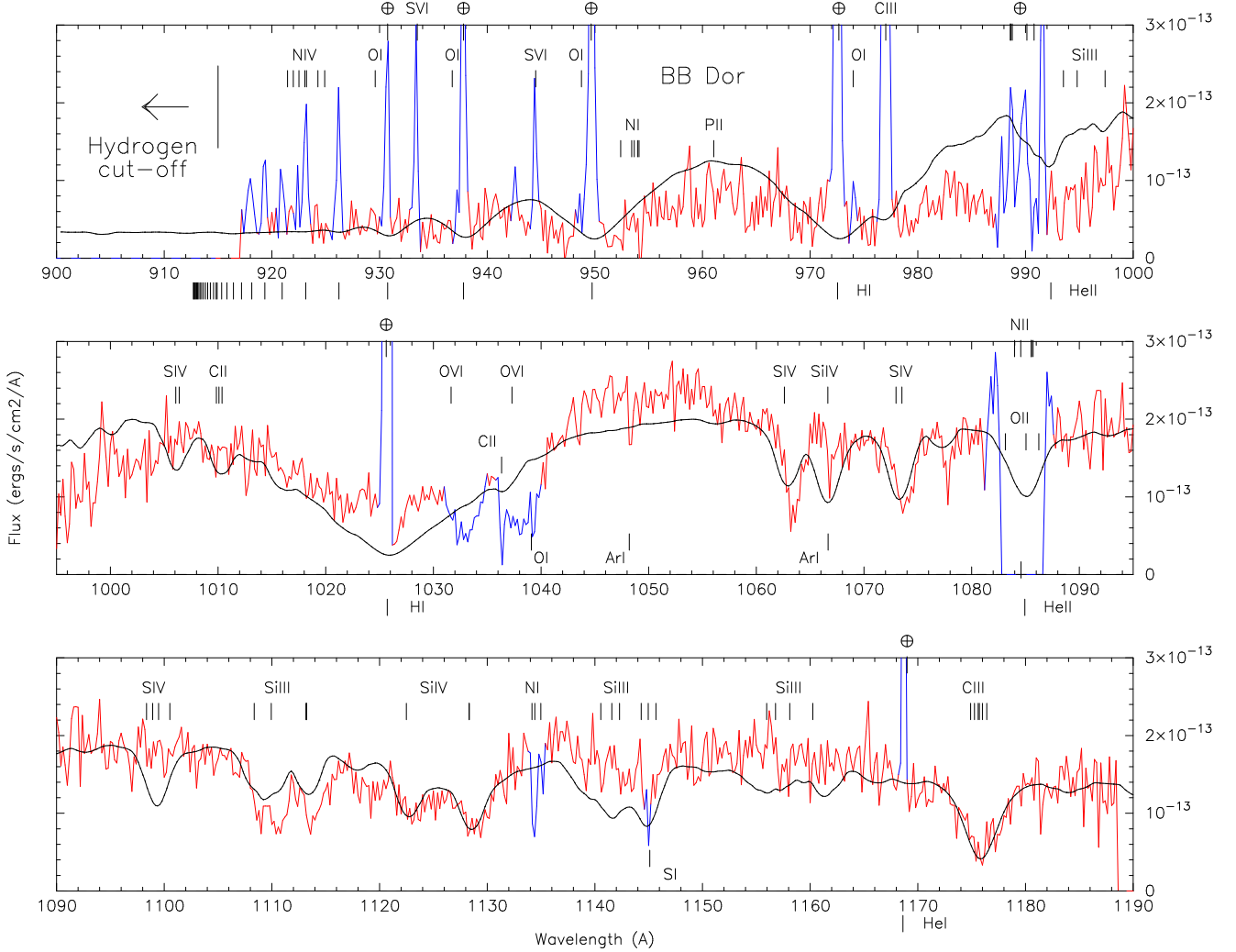}             
\vspace{-5.cm} 
\figcaption{Non-solar abundance single WD model. The best-fit WD model
to the {\it{FUSE}} spectrum of BB Dor is shown where the abundances of
sulfur and silicon have been set to: [S]=20 solar and [Si]=3 solar.
The abundances of all the other elements have been set to solar.  The
WD has a temperature $T=37,000$K, a projected rotational velocity
$V_{rot} \sin{i} =400$km/s, $Log(g)=8.3$. The distance obtained is
$d=$211 pc, and $\chi^2_{\nu}=0.3129$.}
\end{figure}

\clearpage 

\begin{figure}
\plotone{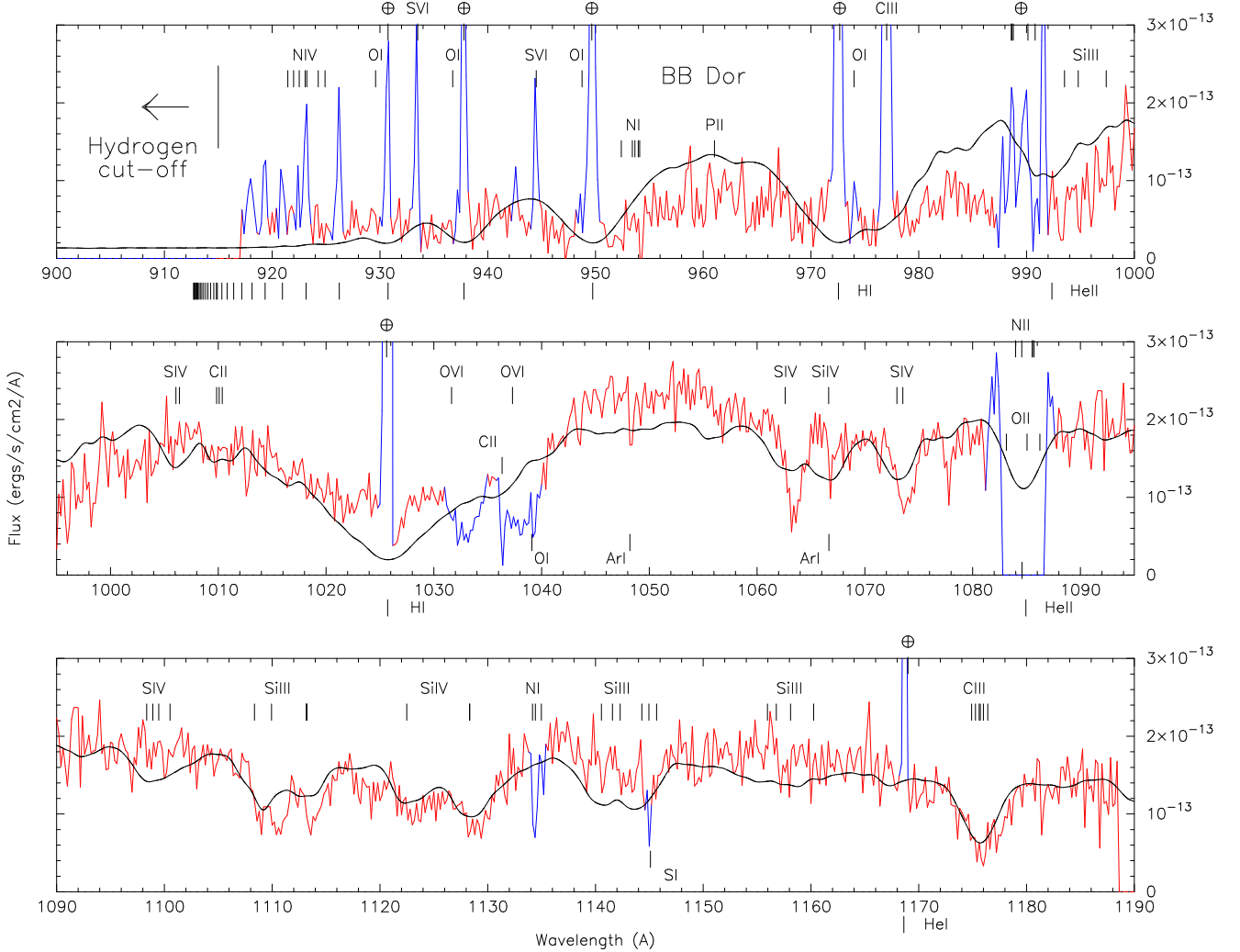}             
\vspace{-5.cm}
\figcaption{Non-solar abundance single disk model.  The {\it{FUSE}}
spectrum of BB Dor is shown together with a one of the best-fit
synthetic disk models.  The model has $M=0.80M_{\odot}$,
$\dot{M}=10^{-9}M_{\odot}$/yr, $i=8^{\circ}$, and gives a distance of
$d=665$ pc and a $\chi^2_{\nu}=0.3000$. The sulfur and silicon
abundances have been set to: [S]=20 solar and [Si]=3 solar. The
abundances of all the other elements have been set to solar.  The low
inclination is needed in order to fit the silicon, sulfur and carbon
absorption lines.}
\end{figure}

\clearpage 

\begin{figure}
\plotone{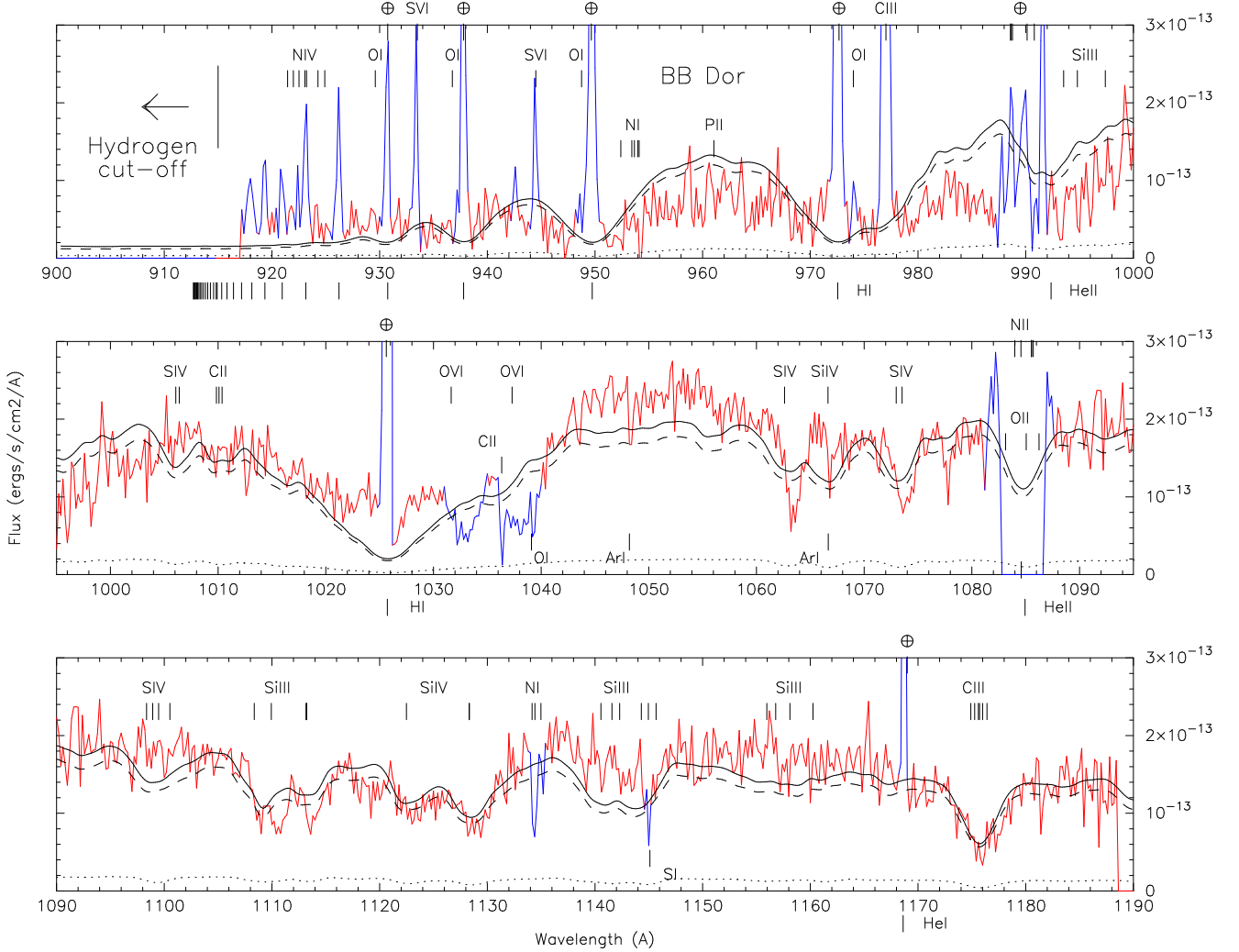}             
\vspace{-5.cm} 
\figcaption{Low-inclination non-solar abundance WD+disk model.  The
{\it{FUSE}} spectrum of BB Dor is shown together with a one of the
best-fit WD+disk models (solid black line), assuming a low inclination
angle. The model has $M=0.80M_{\odot}$, $\dot{M}=10^{-9}M_{\odot}$/yr,
$i=8^{\circ}$, and gives a distance of $d=700$ pc and a
$\chi^2_{\nu}=0.2889$. The sulfur and silicon abundances have been set
to: [S]=20 solar and [Si]=3 solar. The abundances of all the other
elements have been set to solar.  The WD (dotted line) contributes
10\% of the flux while the high-inclination disk (dashed line)
contributes the remaining 90\%.}
\end{figure} 

\clearpage 

\begin{figure}
\plotone{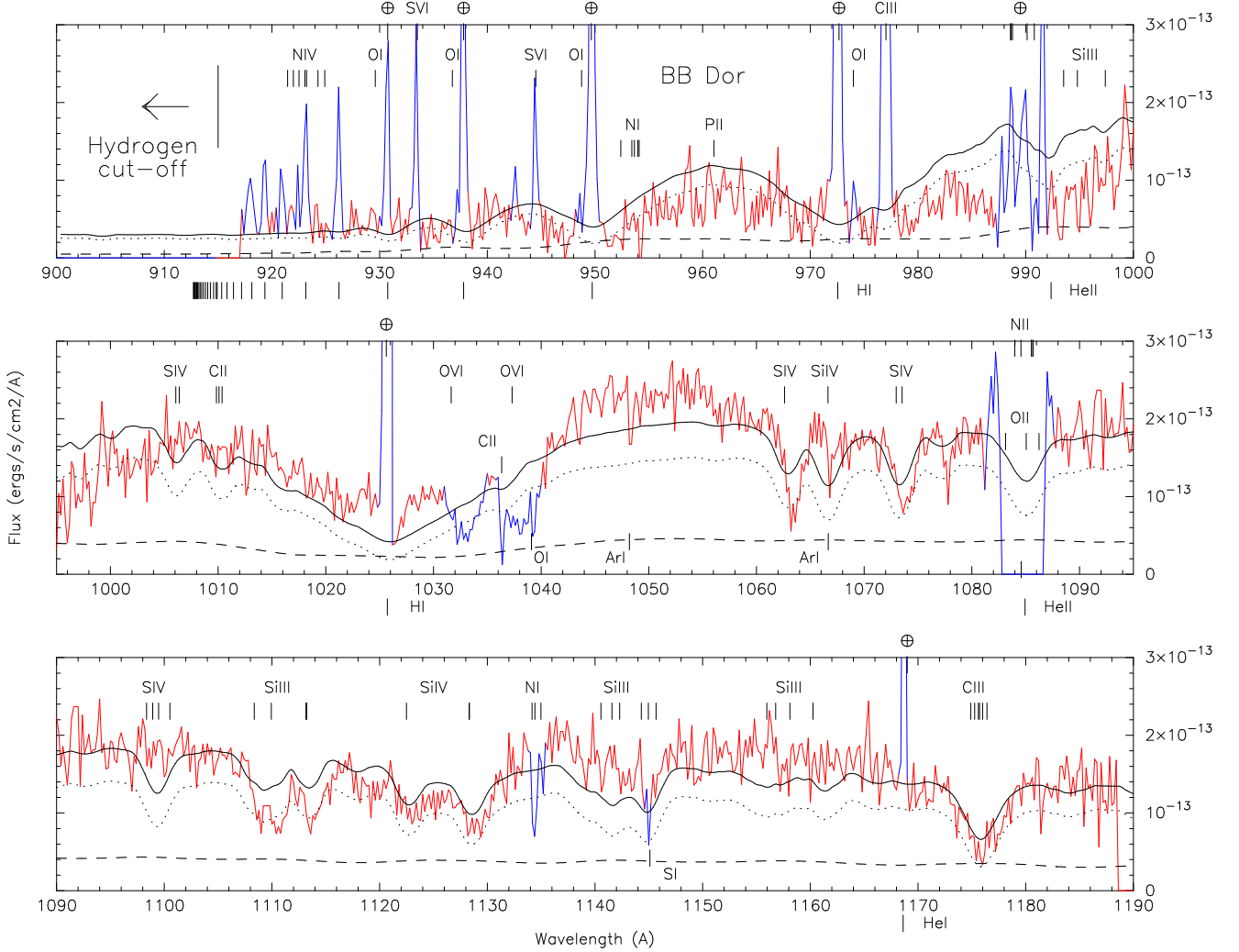}             
\vspace{-5.cm} 
\figcaption{High-inclination non-solar abundance WD+disk model.  The
{\it{FUSE}} spectrum of BB Dor is shown together with a one of the
best-fit WD+disk models (solid black line). The model has
$M=0.80M_{\odot}$, $\dot{M}=10^{-9}M_{\odot}$/yr, and gives a distance
of $d=246$ pc and a $\chi^2_{\nu}=0.2873$. The inclination angle of
the system is kept as a free parameter and the best-fit model gives
$i=80^{\circ}$. The sulfur and silicon abundances have been set to:
[S]=20 solar and [Si]=3 solar. The abundances of all the other
elements have been set to solar.  The fit of the silicon, sulfur and
carbon absorption lines is obtained by having the WD component
dominating the spectrum.  The WD (dotted line) contributes 74\% of the
flux while the high-inclination disk (dashed line) contributes the
remaining 26\%.}
\end{figure}

\clearpage 

\begin{figure}
\plotone{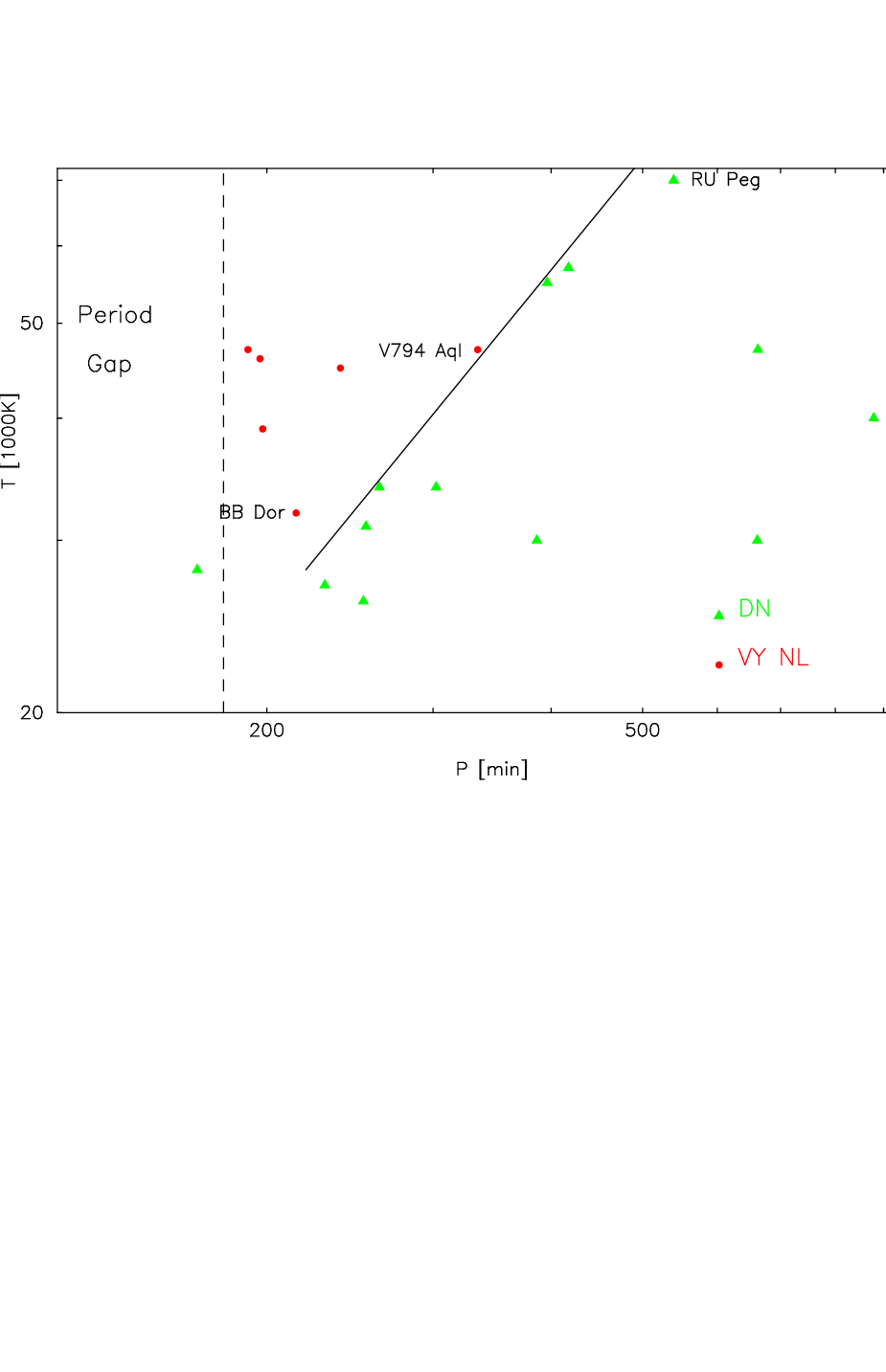}             
\vspace{-10.cm} 
\figcaption{Dwarf nova systems and VY Scl nova-like variables above the 
Period-gap.  The systems have been plotted in the ($T_{eff}$,P)
parameter space on a logarithmic scale. The data for the nova-likes
are taken from \citet{ham08} and references therein; the data for the
dwarf nova systems are taken \citet{sio08} and the references
therein. For RU Peg we have used the more recent estimate from
\citep{god08}. The data point for V794 Aql is an estimate of the WD
temperature based on an IUE spectrum while the system was in a lower
state \citep{god07}. While the data point of BB Dor is lower than the
other VY Scl system there is still a clear separation between the VY
Scl systems and the DN systems as indicated by the slanted line.}
\end{figure} 

\end{document}